\documentclass[%
reprint,
amsmath,amssymb,
aps,
prr,
]{revtex4-2}

\usepackage{graphicx}% Include figure files
\usepackage{dcolumn}% Align table columns on decimal point
\usepackage{bm}% bold math
\usepackage{cancel}
\usepackage{graphicx}
\usepackage{bm}
\usepackage{hyperref}
\usepackage{xcolor}
\newcommand{\sx}{\sigma_x}

\newcommand{\sz}{\sigma_z}
\newcommand{\sI}{\mathbb{I}}
\newcommand{\p}{\partial}

\usepackage[normalem]{ulem}

\newcommand{\sdm}{\varrho}
\newcommand{\Mop}{\mathcal{M}}
\newcommand{\Hop}{\mathcal{H}}
\newcommand{\Tr}{\mathrm{Tr}}

\newcommand{\Prob}{\mathbb{P}}
\newcommand{\Lop}{\mathcal{L}}  % Generator
\newcommand{\dx}{\text{d}}
\newcommand{\inner}[2]{\langle #1|#2\rangle}       % Partial derivative shorthand
\newcommand{\ket}[1]{ |#1\rangle}
\newcommand{\bra}[1]{ \langle#1|}

\def\be{\begin{equation}}
	\def\ee{\end{equation}}
\def\bea{\begin{eqnarray}}
	\def\eea{\end{eqnarray}}

\newcommand{\eqa}[1]{\begin{align}#1\end{align}}%amsmath

\begin{document}
	
	\preprint{APS/123-QED}
	%eno-Anti-Zeno Transition as Motility-Induced Phase Separation in Biased Trajectories
	\title{
		Run-and-Tumble Dynamics and Zeno--Anti-Zeno Transition in Biased Quantum Trajectories}% Force line breaks with \
	
	\author{Aritra Kundu \href{https://orcid.org/0000-0001-7476-8811}{\includegraphics[scale=0.05]{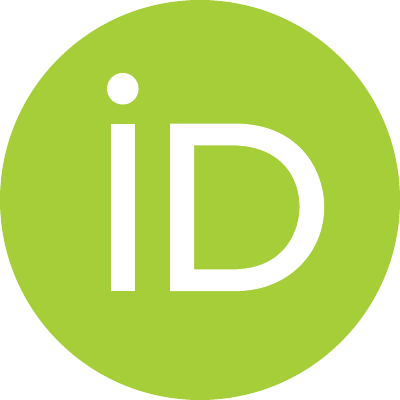}}}
	\affiliation{DPhysMS, University of Luxembourg, \\ 162a, avenue de la Faïencerie, L-1511 Luxembourg}
	\email{aritrakundu@gmail.com,aritra.kundu@uni.lu}%Lines break automatically or can be forced with \
	
	\date{\today}% It is always \today, today,
	
	% --- ABSTRACT (Rigorous & Precise) ---
	\begin{abstract}
		
		We identify the transition from the oscillatory Rabi regime to the localized Zeno/Anti-Zeno regime in continuous measurement and feedback of a qubit as a quantum analogue of Motility-Induced Phase Separation (MIPS).
		A mapping between a biased monitored qubit and a classical ``Run-and-Tumble" active particle is studied. We demonstrate that the competition between coherent Rabi driving (analogous to active motility) and measurement-induced feedback bias (analogous to persistence) mimics the behavior of biological swimmers. This framework provides a picture of measurement-induced phase transitions  using the language of active matter and offering a novel pathway for designing dissipative noisy quantum systems.
	\end{abstract}
	
	\maketitle
	%\tableofcontents
	
	% --- INTRODUCTION ---

	%WHY
	%The study of active matter has revealed that the breaking of detailed balance at the microscopic scale can lead to rich collective phenomena, such as motility-induced phase separation and ratchet effects \cite{ramaswamy_mechanics_2010,malakar_steady_2018}. 
	
	Active matter systems, collections of entities that consume energy to propel themselves, have become a cornerstone of non-equilibrium statistical physics. A paradigmatic model in this field is the Run-and-Tumble Particle (RTP) in one dimension, where a particle undergoes persistent motion ("run") with a velocity that switches stochastically between discrete states \cite{ramaswamy_mechanics_2010,malakar_steady_2018}. By breaking microscopic detailed balance, RTPs give rise to rich collective phenomena that have no equilibrium counterpart, such as ratchet effects and motility-induced phase separation (MIPS), where self-propulsion paradoxically leads to particle trapping. While these behaviors are traditionally associated with biological systems occurring on macro-mesoscopic timescales \cite{tailleur_sedimentation_2009,cates_motilityinduced_2015,tailleur_sedimentation_2009,maes_dynamical_2017,bruyne_survival_2021}, the fundamental physics of "activity"-specifically the interplay between persistence and stochasticity is universal. Unlike classical quantum-trajectory studies, our active-matter lens provides an intuitive framework for understanding the interplay of measurement bias and feedback in quantum evolutions, highlighting the emergent persistent behavior at scales where competing explanations often fall short in addressing the transitions between observable states.
	\begin{figure}
		\includegraphics[scale=0.34]{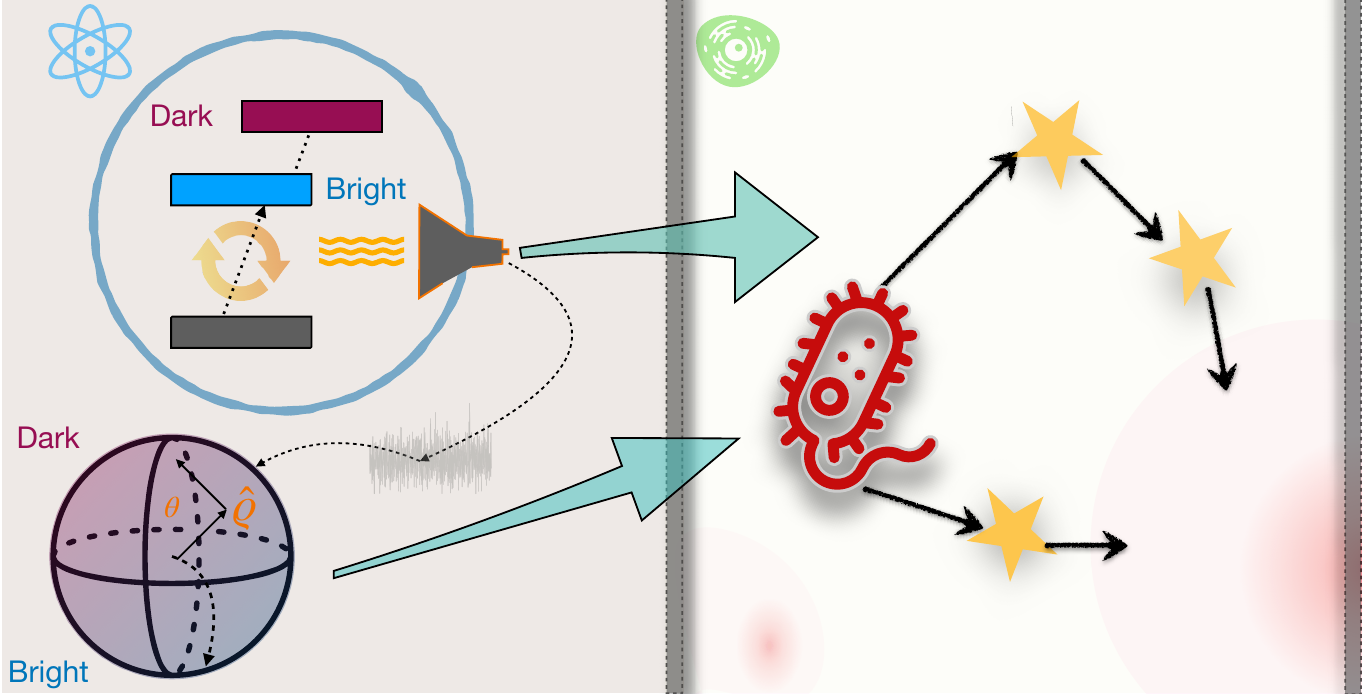}
		\caption{\textbf{Analogous stochastic pathways  driven by noisy measurement records.}
			\textbf{LEFT}, Quantum state evolution under continuous monitoring. A noisy two-level  quantum system possessing a radiatively active ``bright" state and a metastable ``dark" state is coupled via a Rabi drive. A detector continuously monitors the qubit, generating a stochastic sequence of ``clicks". This filtered, noisy measurement record conditions the evolution of the system's state estimate, depicted as a stochastic trajectory of the density matrix $\hat \varrho$ on the Bloch sphere, which evolves toward the dark state due to measurement backaction. The bias can be due to selective weighting of the quantum trajectories. Post-processing normalization maintains the system to be in pure state at single trajectory level.
			\textbf{RIGHT}, Bacterial navigation via  run-and-tumble dynamics confined between two walls. A bacterium explores a nutrient gradient surrounding a symmetrically placed ``food source" (red spots). Its locomotion consists of straight swimming intervals (``runs") punctuated by stochastic reorientation events (``tumbles") in a fluctuating environment. This noisy sensory input regulates the internal switching probability between running and tumbling states, biasing the bacterium's random walk. The quantum jumps in the atom can be viewed through the lens of "tumbling" of the bacterial motion.}
		
		\label{fig:comp}
	\end{figure}
	Concurrently, in the quantum domain, continuous measurement theory allows us to track the evolution of a single quantum system in real time. These trajectories evolve stochastically, driven not by thermal fluctuations, but by the backaction of the measurement feedback \cite{wiseman_quantum_1993}. Recent technological advancements in superconducting circuits have enabled the direct observation of these dynamics, including non-Hermitian evolution \cite{murch_observing_2013} and incomplete quantum jumps, known as ``spikes" \cite{minev_catching_2019}. The statistics of these trajectories, specifically the distribution of first-passage times \cite{dhar_quantum_2015,yin_restart_2025,yin_resonances_2025,muga_quantum_1999}, and the structure of spikes in diffusive \cite{tilloy_spikes_2015,bernardin_spiking_2023,benoist_emergence_2021,benoist_large_2014,horowitz_quantumtrajectory_2012,walter_field_2022} or Poissonian limits \cite{sherry_spikes_2025}, suggest a hidden active character in monitored quantum evolution.
	
	In this Letter, we bridge these two disparate worlds. We investigate the stochastic dynamics of a qubit subjected to simultaneous coherent driving and continuous measurement with feedback. We propose that the Zitterbewegung-like oscillations of the quantum state are physically equivalent to the motion of a classical active particle. Fig.\eqref{fig:comp} shows a schematic of the microscopic noisy three-level system being continuously monitored and driven by a Rabi drive. Reconstructing the time-dependent state of the system ($\ket{\psi}$) involves post-processing a stochastic signal from the detector, where durations of a rapid increase in the number of clicks ("Bright" states) are punctuated by no clicks ("Dark" states). A biased trajectory involves post-processing the signal selectively. Specifically, we show that the qubit’s trajectory on the Bloch sphere coarse-grains to the dynamics of a persistent Run-and-Tumble Particle confined in a bounded domain. A surprising connection emerges: second-scale biological motion sheds light on sub-microsecond quantum evolution, revealing that complex quantum phenomena, such as the Zeno-anti-Zeno transition, can be reinterpreted as manifestations of motility-induced trapping. Moreover, we anticipate that for a measurement strength along role of enviromental noise is weaker than a critical value, namely $\gamma_{eff}^* \sim O(1)$, the Zeno-Anti-Zeno transition disappears. To the best of my knowledge, this was not explicitly reported before, setting the stage for later illustration in the phase diagram.
	
	The significance of this mapping is twofold. First, it allows us to import the extensive analytical toolbox of classical active matter to characterize the statistics of the quantum trajectory, providing solutions that are otherwise difficult to derive. Second, it offers a novel physical interpretation of the Quantum Zeno effect in terms of active transport: the ``freezing" of the quantum state is revealed to be dynamically equivalent to the trapping of an active particle, a phenomenon that occurs precisely when the drift velocity exceeds the tumbling rate.
	
	\begin{figure*}[htbp]
		\centering
		
		\includegraphics[width=\linewidth]{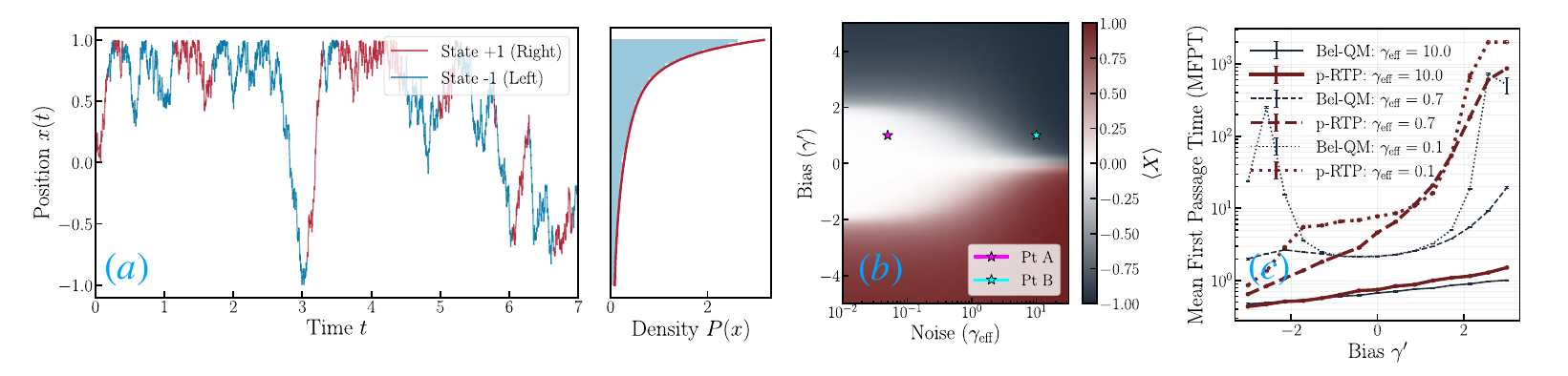}
		\caption{\textbf{Classical-Quantum connection} 
			(a) Single trajectory of the effective coordinate $x(t)$ of p-RTP model from Eq.\eqref{eq:sde_p-RTP}. The red-blue markings reflect the current internal state. Notice that the current internal state does not always mimic the instantaneous direction of the particle. The right panel shows the stationary distribution, as can be verified from the exact solution given in the main text.
			(b) Phase diagram showing the transition from Zeno (Maroon) to Anti-Zeno (Dark Grey) stationary distributions obtained from \textit{stochastic averaged-} Eq. \eqref{eq:BBeqn} (Avg-QM). The effect of the feedback bias $\gamma'$ acts as a symmetry-breaking drift field.
			(d) The Mean First Passage Time (MFPT) to the boundary. The maroon lines represent the prediction from the p-RTP model, while black lines represent numerical simulations of the quantum Belavkin SME: Eq.\eqref{eq:BBeqn} (Bel-QM). The agreement in the high-noise limit validates the effective p-RTP description.
		}
		\label{fig:maxwell_demon_trajectory}
	\end{figure*}
	%VARIABLE DEFS
	
	In what follows, we first derive the explicit steady-state solutions for the classical p-RTP model in a bounded domain $x \in [-\ell,\ell]$. We then discuss the quantum counterpart, detailing the dynamics of a continuously monitored two-level quantum system (qubit) subject to measurement-based feedback. To rigorously validate the connection between these systems, we derive the characteristic equation governing both the classical steady-state distribution of the Fokker-Planck variable $P(x)$ and the noise averaged Liouvillian superoperator evolved quantum state at large time $\hat \rho = \langle \ket{\psi}\bra{\psi} \rangle $ of the biased feedback-controlled quantum trajectories. We demonstrate that the spectral function for both systems reduces to an identical cubic equation in Weierstrass form:\eqa{t^3 + pt + q = 0,
		\label{eq:root_eq}}where the physical requirement for a valid probability density implies that the roots must be real. This condition is strictly satisfied when the discriminant is negative, $\Delta_t = -4p^3 - 27q^2 < 0$, ensuring the stability of the non-equilibrium steady state in the presence of reflective boundaries.
	
	%–
	
	% --- SECTION II ---
	\section{Classical Model: }

	\textit{Persistent RTP in a Bounded Domain.} We consider a classical active particle confined to a one-dimensional domain $x \in [-\ell, \ell]$, whose dynamics are governed by the Langevin equation:\begin{equation} d{x} = [v_d + s(t) v]dt + \sqrt{2D}dW_t. \label{eq:sde_p-RTP}\end{equation}Here, $v_d$ represents a constant drift velocity, $v$ is the active speed, and $D$ is the diffusion coefficient associated with the standard Wiener process $W_t$ (satisfying the Itô rule $dW_t^2 = dt$). The term $s(t) \in {-1, 1}$ represents a telegraph noise that flips between states at a rate $\lambda$. As shown in Fig. \ref{fig:maxwell_demon_trajectory}(a), the instantaneous internal state $s(t)$ dictates the local direction of the particle’s active motion. To determine the long-time statistics of this system, we solve the corresponding stationary Fokker-Planck equation subject to zero-flux reflective boundary conditions \cite{malakar_steady_2018}, which yields the steady-state probability distribution $P(x)$.
	%–
	
	\textit{Characteristic Equation and Explicit solution of steady-state:}
	
	The piling up of the probability density at the boundaries is reminiscent of non-hermitian skin effects \cite{roccati_nonhermitian_2021}.
	The exact solution presented before matches well with the numerical simulation of the Langevin equation in Fig.\ref{fig:maxwell_demon_trajectory}(a) (red line).
	The steady-state probability density $P(x)$
	is a superposition of exponential modes, $e^{(t+ \frac{2v_d}{3D}) x}$ where $t$ satisfys  Eq.\eqref{eq:root_eq}
	where the coefficients  $p$ and $q$ are both functions of $v,v_d,D,\lambda$  \footnote{See Appendix. To be published later: $p = \beta - \frac{\alpha^2}{3}$ and
		$q = \frac{2\alpha^3}{27} - \frac{\alpha\beta}{3} + \delta$
		with $\alpha = - 2\frac{v_d}{D}$, $\beta = \frac{1}{D^2}(v_d^2 - v^2 - 2D\lambda)$
		and $\delta = \frac{2\lambda }{D^2}v_d $}
	\cite{malakar_steady_2018}.
	
	\textit{Asymptotic results:}\label{sec:asym}
	\begin{figure*}
		\centering
		\includegraphics[width=1.0\linewidth]{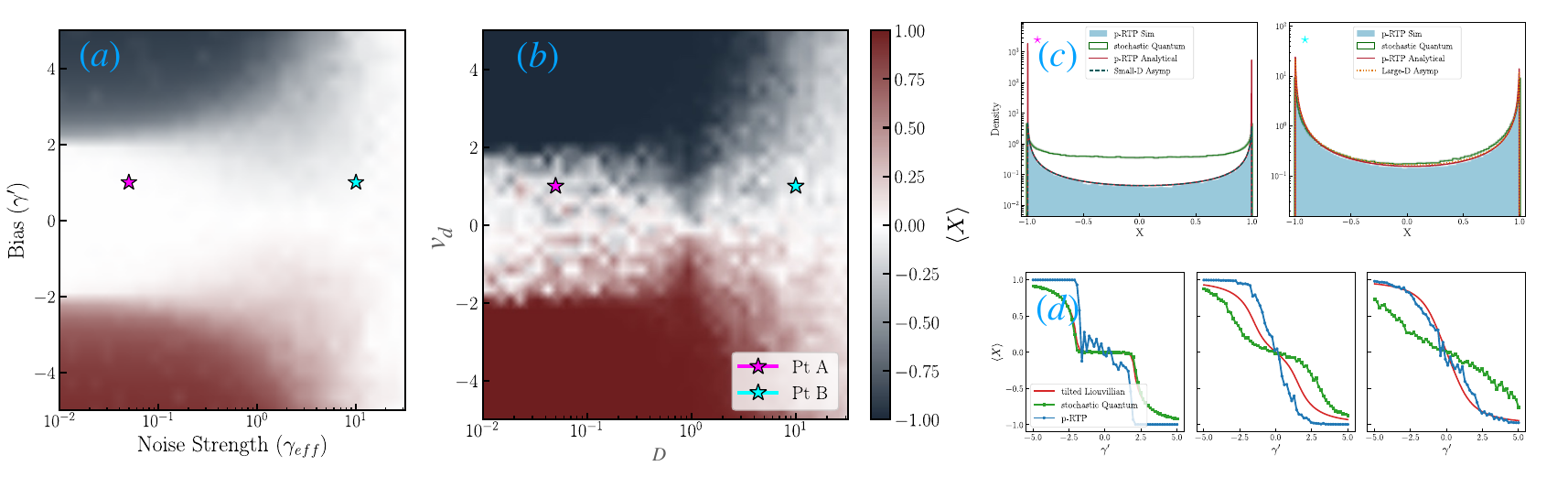}
		\caption{\textbf{ Measurement-induced transitions as motility-induced phase transition} (a) Phase diagrams displaying the mean position $\langle X \rangle$ as a function of the effective diffusion rate $\gamma_{\text{eff}}$ and bias parameter $\gamma'$. computed from Bel-QM (b). Shows the averaged mean position from the effective p-RTP SDE approximation. The stars mark two specific parameter regimes: Point A (magenta, low diffusion $\gamma_{\text{eff}}=v_d= 0.05$) and Point B (cyan, high diffusion $\gamma_{\text{eff}}=v=10.0$). (c) Steady-state probability density distributions $P(X)$ for the parameters corresponding to points A and B for the p-RTP model and the quantum simulation (Bel-QM). The plots compare the p-RTP simulation (blue histogram), the quantum simulation (green line), the exact solution (red line), and the small-diffusion asymptotic approximation (dashed teal line) as formulas given in the main text. The statistics of the phase becomes accurate in the higher diffusion limit where the quantum effects has been washed out as also evidence from the first passage times Fig. \ref{fig:maxwell_demon_trajectory}(c) (d) One-dimensional slices of the phase diagram at fixed diffusion levels ($\gamma_{\text{eff}} = 0.02, 0.7, 2.0$). The curves compare the mean position $\langle X \rangle$ obtained from the tilted Liouvillian evolution (Avg-QM) (purple), the stochastic quantum simulation (Bel-QM) (green), and the hyperbolic transformed p-RTP model Eq.\eqref{eq:RTP} with $\ell = \pi$ (blue). }
		\label{fig:fullphase}
	\end{figure*} To gain physical insight into the system's phase behavior, we derive explicit results for the limiting cases of low ($D \to 0$) and high ($D \to \infty$) diffusion. In the low-diffusion limit, the steady-state density $P(x)$ describes an asymptotic theory combining a smooth exponential bulk with significant accumulation at the boundaries:\begin{eqnarray}P(x) \propto e^{r_1 x} + m_{\delta}^{\pm} \delta(x \pm \ell).\end{eqnarray}Here, the bulk decay rate is given by $r_1 = \frac{2\lambda v_d}{v^2 - v_d^2}$, while $m_{\delta}^{\pm}$ represents the finite $\mathcal{O}(1)$ probability mass accumulated at the boundaries $x = \pm \ell$:$$m_{\delta}^{\pm} = \frac{1}{2} \frac{v_d}{v} \frac{e^{\pm r_1 \ell}}{\sinh(r_1 \ell) + \frac{v_d}{v}\cosh(r_1 \ell)}.$$Physically, this result generalizes the phenomenon of active-particle sedimentation in bounded domains \cite{tailleur_sedimentation_2009,cates_diffusive_2012}. We observe that the characteristic length scale diverges when the drift velocity matches the active speed ($v_d = v$), marking a critical transition in the trapping efficiency. Conversely, in the high-diffusion limit ($D \to \infty$), the telegraph noise flipping becomes irrelevant compared to the thermal fluctuations. The steady-state density simplifies to a linear profile which is independent of flipping velocity:$$P(x) \approx \frac{1}{2\ell} \left( 1 + \frac{v_d}{D} x \right).$$With these classical baselines established, we next turn to the quantum setup of a continuously monitored two-level system and demonstrate its mapping to this p-RTP model.
	
	%--
	
	% --- SECTION III ---
	\section{Quantum Model: }
	\textit{The Monitored Qubit with Conditional Feedback.} We consider a two-level system (qubit) subjected to a coherent Rabi drive of strength $J$, described by the Hamiltonian $\hat{H}_0 = J\sx$, generating rotations around the $x$-axis. Simultaneously, the system is continuously monitored via the projection operator $\hat{L} = \sigma^-\sigma^+$ which induces pure dephasing in the $z$-basis. Crucially, because the measurement operator does not commute with the Hamiltonian ($[\hat L,\hat H_0] \neq 0$), the measurement is not Quantum Nondemolition (QND). Instead, the monitoring exerts a stochastic backaction on the state $\ket{\psi}$, creating a competition between the coherent drive and the measurement-induced collapse.This configuration constitutes a canonical model for open quantum systems. 
	Physically, the Hamiltonian ($H = J\sigma_x$) generates rotations around the $X$-axis.  The measurement process acts as dephasing in the $Z$-basis, which destroys any coherence between the ``up" and ``down" states and we expect a zero average value for the measurement for low enough measurement strength (Pt. A in Fig. \ref{fig:maxwell_demon_trajectory}(b)). {When the measurent bias is very large (the Zeno limit), the system freezes in the eigenspace of the measurement operator (the states $|0\rangle = (1,0)^{\mathsf T}$ (Bright) or 
		$|1\rangle = (0,1)^{\mathsf T}$ (Dark))
		.  A small bias results in the state to be  $\ket{\psi}_{\pm} \approx \frac{1}{2}(\ket{0}\pm \ket{1})$ where the sign $\pm$ denotes Zeno freezing  or acceleration respectively (Pt. B in Fig. \ref{fig:maxwell_demon_trajectory}(b))\footnote{The system in $\ket{\psi}_{pm} $}.}
	
	Previous studies have analyzed distinct regimes of this dynamics, including diffusive trajectories \cite{martinez-azcona_quantum_2025}, Poissonian ``spiking" behavior \cite{sherry_spikes_2025}, and Quantum Zeno dynamics \cite{kumar_quantum_2020,dubey_quantum_2021,dubey_quantum_2023}. While this setup has been realized experimentally using superconducting circuits \cite{murch_observing_2013}, such experiments were often limited by no-jump post-selection. Here, we analyze the full stochastic evolution without such restrictions \footnote{To match the specific physical system introduced in \cite{martinez-azcona_quantum_2025} in the context of a stochastic non-Hermitian Hamiltonian, we perform the parameter mapping: $\gamma_{\text{eff}} = \gamma \Gamma_e^2$ and $ \gamma’ = -\Gamma_e + 2\gamma \Gamma_e^2$}.

	%-----
	The dynamics of the monitored quantum system are described by the tilted Belavkin-Barchielli-Zakai equation (Bel-QM) \cite{barchielli_quantum_2009, belavkin_new_1989,_quantum_d, chetrite_nonequilibrium_2015,perfetto_thermodynamics_2022a}. This stochastic differential equation governs the evolution of the system's density matrix $\hat{\sdm} = \ket{\psi}\bra{\psi}$, representing the observer's ``best estimate" of the state conditioned on the biased measurement record collected up to time $t$ (See Appendix):
	\begin{equation}d\hat{\sdm} = \mathcal{L}^{\gamma_{\text{eff}}}[\hat{\sdm}] dt + \mathcal{I}^{\gamma_{\text{eff}},\gamma'}[\hat{\sdm}]. \label{eq:BBeqn}\end{equation}The first term represents the average dissipative evolution, described by the standard Lindblad superoperator:\begin{equation}\mathcal{L}^{\gamma_{\text{eff}}}[\hat{\sdm}] = -i[\hat{H}_0,\hat{\sdm}] dt + \gamma_{\text{eff}}\left(2\hat{L}\hat{\sdm}\hat{L}^\dagger - \{\hat{L}^\dagger\hat{L},\hat{\sdm}\}\right) dt,\end{equation}where $\gamma_{\text{eff}}$ is the effective diffusion constant arising from the measurement back-action. The second term is the stochastic ``innovation," which updates the state based on the fluctuations in the measurement stream:\begin{equation}\mathcal{I}^{\gamma_{\text{eff}},\gamma'}[\hat{\sdm}] = \left(\gamma' dt - \sqrt{2\gamma_{\text{eff}}} dW_t\right)\left(\hat{L}\hat{\sdm} + \hat{\sdm}\hat{L}^\dagger  - \text{Tr}[\hat{\sdm}(\hat{L} + \hat{L}^\dagger)]\right).\end{equation}Physically, the parameter $\gamma'$ quantifies the ``information" or bias collected by the observer; mathematically, it represents the deviation of the measurement record from the standard martingale assumptions of unbiased noise. Following Ito stochastic calculus rules, one can show explicitely that the above dynamics preserves purity of the state. It is crucial to distinguish our approach from alternative methods. We employ single-trajectory normalization, a technique routinely used in quantum trajectory theory \cite{dubey_quantum_2021, dubey_quantum_2023, benoist_emergence_2021, cornelius_spectral_2022}, where trace preservation is enforced at every time step of the stochastic evolution. This contrasts with the ``linear" approach (e.g., used in \cite{martinez-azcona_quantum_2025}, where one averages over the noise before imposing trace preservation. As these two approaches generally yield different dynamics in non-linear feedback scenarios, our choice highlights the importance of preserving the physical validity of the single-shot quantum state.

	To isolate the deterministic component of the evolution, we first consider the zero-noise limit ($dW_t \to 0$). In this regime, the dynamics are governed solely by the averaged Liouvillian superoperator, $\mathcal{L}^{\gamma_{\text{eff}}}$, acting on the average density matrix $\hat \rho = \langle \hat{\varrho} \rangle$.Using the operator basis $E_{jk} \equiv |j\rangle\langle k|$ (where $\inner{k}{j} = \delta_{kj}$), this superoperator can be written in a compact  form:\begin{equation}\mathcal{L}^{\gamma_{\text{eff}}} = \sum_{j=1}^4 d_j E_{jj} + \Big[i\big(E_{12}+E_{34}-E_{13}-E_{24}\big)+\text{h.c.}\Big], \label{eq:Liovillian}\end{equation}where the diagonal decay rates are given by $d_1=0$, $d_2=d_3=-\gamma_{\text{eff}}/2$, and $d_4=-\gamma'$.The spectral properties of this operator determine the system's long-time behavior. We find that the characteristic polynomial governing the steady-state eigenvalues corresponds exactly to the cubic Weierstrass form introduced in Eq. \eqref{eq:root_eq}\footnote{Here, the coefficients are derived explicitly as $p = 4 - \frac{3\gamma_{\mathrm{eff}}^{2}-3\gamma_{\mathrm{eff}}\gamma’+\gamma’^{2}}{12}$ and $q = -\gamma_{\mathrm{eff}} +\gamma’\frac{\gamma_{\mathrm{eff}}^{2}-\gamma_{\mathrm{eff}}\gamma’+8}{24}$.} The bias parameter $\gamma'$ plays a dual role in determining the nature of the quantum trajectory. In the unbiased limit ($\gamma' = 0$), the innovation term exactly counterbalances the decoherence induced by the Lindblad dissipation. This cancellation preserves the purity of the state, characteristic of standard quantum state diffusion. In the persistant limit ($\gamma' \neq 0$), when the measurement is biased, the innovation term no longer merely filters noise. Instead, it actively biases the fluctuations, generating a net ``propulsion" effect.  This force prevents the system from settling into a trivial equilibrium, effectively driving the ``active" motion mapped to the p-RTP model.
	\section{Information geometric mapping}
	\textit{Mapping Information to Propulsion.} To bridge the quantum and classical descriptions, we perform a nonlinear coordinate transformation on the Bloch sphere dynamics  \cite{zyczkowski_geometry_2006}. We define the variable $\zeta_t = \operatorname{arctanh}(X_t)$, where $X_t = \langle \sigma_z \rangle_t$ represents the quantum expectation value of the Pauli-$Z$ operator measuring populations. Through the Gudermannian relation, $2\arctan(X_t) = \operatorname{gd}(2\zeta_t)$, the bounded dynamics of the quantum state ($X_t \in [-1,1]$) are mapped to an unbounded domain ($\zeta_t \in (-\infty,\infty)$). In the high-diffusion limit, the effective dynamics of this transformed coordinate is governed by the following Stochastic Differential Equation (SDE):\begin{equation}dx_t \approx [-\gamma’ + 2Js(t)] dt + \sqrt{2\gamma_{\text{eff}}} dW_t. \label{eq:RTP}\end{equation}Here, the bias parameter $\gamma'$ acts as a constant drift, while the coherent Rabi drive $J$ is reinterpreted as an active propulsion term. The validity of the active nature of the biased quantum trajectory in Eq.~\eqref{eq:RTP} rests on a crucial physical insight: in the large-noise limit, the coherent Rabi oscillation is effectively ``chopped" into instantaneous switching events. However this is not the only contribution, as the noise itself induces some flipping events. Mathematically, this approximates the non-Markovian, state-dependent flip rate ($\cosh(x)$). One can approximate this as Markovian telegraph process $s(t) \in \{-1, 1\}$, with a constant switching rate given by:$$\lambda(\gamma_{\text{eff}}, J) \approx \frac{J(4/\ell - 1)}{\log(\gamma_{\text{eff}})}.$$This approximation renders the quantum drive statistically equivalent to the stochastic tumbling of an active particle. Consequently, as the noise increases, the run-and-tumble picture becomes exact, and the full quantum solution converges to the p-RTP prediction. To maintain correspondence with the bounded p-RTP model and ensure numerical stability, we simulate the transformed variable truncated to the domain $x_t \in [-\ell, \ell]$. We compute the steady-state quantum expectation value by averaging the back-transformed variable over long-time simulations: $\langle X_t \rangle_{ss} = \langle \tanh(x_t) \rangle_{ss}$. As shown in Fig.~\ref{fig:maxwell_demon_trajectory}, the First-Passage Time statistics confirm this dynamical equivalence, demonstrating that the ``active" character of the trajectories emerges naturally from the competition between measurement and drive.
	
	\section{Results}
	\textit{{NESS Zeno--Anti-Zeno Phase Diagram}:}
	\label{sec:zeno}
	We now analyze the non-equilibrium steady-state (NESS) properties of the system. As illustrated in Fig.\ref{fig:fullphase}(a), the feedback-controlled qubit realizes a rich phase diagram characterized by a Zeno--anti-Zeno crossover. The Zeno transition initially identified by Misra and Sudarshan \cite{ sudarshan_interaction_1976,misra_zenos_1977}. Contrary to the stabilization of states, frequent measurements can also accelerate decay—the Anti-Zeno Effect (AZE). In line with this, Balachandran and Roy established an exact differential equation for continuous time-dependent measurements, proving that for certain conditions for evolution, the "watched kettle" is sure to boil, establishing a quantum anti-Zeno paradox \cite{balachandran_quantum_2000a,home_conceptual_1997,greenfield_unified_2025,streed_continuous_2006}. The quantum Zeno effect (QZE), predicting the inhibition of quantum evolution under frequent observation, has evolved from a foundational paradox \cite{allcock_time_1969,echanobe_disclosing_2008,streed_continuous_2006,home_conceptual_1997,dhar_preserving_2006} into a verifiable physical phenomenon and a resource for quantum technology \cite{greenfield_unified_2025,annby-andersson_maxwells_2024,mukherjee_antizeno_2020,boyanovsky_quantum_2019,patsch_quantum_2020}. Here, we explore how this transition emerges from the interplay between continuous measurement bias and dissipation. Typical experimental setups using superconducting qubits with a measurement latency speed in less than $200 $ ns allow these phenomena to be observed, as they fall squarely within the predicted Rabi regime outlined in this study. Concrete lab parameters like these underscore the experimental viability of our theoretical framework and offer a compelling invitation to experimentalists to test these predictions.
	
	\textit{The Zeno Regime (Localization).---}
	In the regime of significant positive bias ($\gamma' \gg J$), the feedback mechanism (acting as a Maxwell's demon) strongly favors the subspace aligned with the bias. Consequently, the dynamics enter a \emph{Zeno-like freezing} regime: transitions are suppressed, and the system becomes localized near the boundary $\langle X\rangle \approx -1$. This localization is consistent with the ``measurement-modified state selection'' predicted in Refs.~\cite{kofman_acceleration_2000}. Symmetrically, a significant negative bias stabilizes the opposite sector, locking the system to $\langle X\rangle \approx +1$ suggesting Zeno-acceleration.
	
	\textit{The Rabi (Delocalization).---}
	Between these localized domains, we observe a broad \emph{Rabi regime}. Here, neither the bias $\gamma'$ nor the effective dissipation $\gamma_{\mathrm{eff}}$ is sufficient to arrest the dynamics. Instead, the stochastic back-action from the feedback effectively competes with the deterministic bias, manifesting as an effective ``activity.'' These enhanced fluctuations drive the switching rate, forcing the system toward a delocalized steady state with vanishing mean real coherence ($\langle X\rangle \approx 0$) with switching between the dark and bright states.
	
	\textit{The Phase Boundary.--- }
	The transition between these regimes is defined by the competition between \emph{deterministic dissipation} (which tends to freeze the state) and \emph{stochastic biasing} (which tends to drive it).
	
	{For weak dissipation ($\gamma_{\mathrm{eff}} \lesssim 1$),} the crossover to the localized phase occurs approximately when the bias exceeds the coherent drive scale, $|\gamma'| \simeq 2J$.
	
	{While in the strong feedback regime,} as $\gamma’$ increases, the enhanced fluctuations counteract the dissipative freezing. Consequently, substantially larger bias is required to maintain the Zeno subspace, shifting the transition to larger values of $\gamma’$.
	
	\textit{Validation of the Classical Active Matter Mapping.---}
	Comparing the quantum phase diagram [Fig.2(a)] with the classical p-RTP model [Fig.2(b)] reveals that the classical active matter description captures the essential features of the quantum simulation.% at the average level.
	\emph{High-Diffusion Limit:} As the diffusion constant increases, the ``active'' picture becomes accurate. The classical p-RTP solution for $\langle X_t \rangle$ matches the averaged Lindblad equation, and the full probability distributions overlap to a good degree (within $\sim 2\%$ error, see Fig.~\ref{fig:fullphase}(c), Point B).
	\emph{Small-Diffusion Limit:} In the low-noise regime, the quantum-to-classical map correctly identifies the phase transition boundaries at $\gamma' = \pm 2J$. However, closer inspection of the full distribution [Fig.~\ref{fig:fullphase}(d)] reveals deviations due to quantum-classical transition approximations. While the mean values agree qualitatively, the full quantum distribution displays structure not captured by the simple Langevin description, highlighting the emergence of uniquely quantum correlations in the low-noise limit.
	
	% --- CONCLUSION ---
	
	In this Letter, we have investigated the active dynamics of biased continuous quantum measurements using stochastic processes with telegraphic noise. We established an approximate mapping between a continuously monitored qubit under feedback and a classical persistent Run-and-Tumble Particle (p-RTP) confined to a bounded domain. Looking forward, this mapping presents an exciting potential for extension to more complex systems, such as multi-qubit chains or higher-dimensional active particles. Exploring these directions could provide deeper insights into the broader applicability of our framework and stimulate future research dialogues in both quantum mechanics and active matter fields.
	
	This correspondence provides a mechanism for understanding measurement-induced transport: the Quantum Zeno-Anti-Zeno transition of the qubit is revealed to be analogous to the 'motility-induced shape transition' observed in classical active matter when viewed in the correct lens. This framework offers a physical interpretation of the Quantum Zeno effect as 'motility-induced trapping,' where the suppression of coherent motion by measurement is dynamically equivalent to the trapping of an active particle by confinement. Conversely, the phenomenon of 'motility-induced trapping' within active matter could be explored through the dynamics of qubits, highlighting the interdisciplinary utility of this mapping.
	
	Beyond conceptual unity, our results demonstrate that powerful tools from nonequilibrium statistical mechanics, in particular, large-deviation theory and current-fluctuation analysis for RTPs, provide a natural framework for characterizing the statistics of quantum trajectories. Extending this approach to study full-time dynamics and including non-Markovian flipping rates to capture the complete quantum evolution remains a promising avenue for future work.

	Crucially, this proposal is not merely abstract; the system is readily implementable with existing superconducting qubit technologies \cite{murch_observing_2013,minev_catching_2019}, allowing for experimental verification of these active matter analogies in the quantum domain. A projective measurement discussed here can also be readily implemented in a quantum computer making the predictions verifiable using current technologies. We note that while our two-level system with measurement bias exhibits structural similarities to not completely positive maps recently applied to continuous-variable systems \cite{antonov_engineering_2025,antonov_modeling_2025}, our framework extends these concepts to the discrete dynamics of qubits and studying the full distribution giving detailed information of the quantum state in specific limits.
	
	\begin{acknowledgments}
		I thank Raphael Chetrite, Cedric Bernardin, Abhishek Dhar, Adolfo Del Campo, Aurelia Chenu, Raphael Wittowiski, Pablo Martinez Azcona, and   Komal Kumari for valuable discussions on related topics.  I thank Shiue-Yuan-Shiau and Oskar Prosniak for improving the quality of the document. This work is supported by FNR CORE junior grant no. 17132054. This research was supported in part by the International Centre for Theoretical Sciences (ICTS) for the program Quantum Trajectories (code: ICTS/qt2025/01).
	\end{acknowledgments}
	% --- BIBLIOGRAPHY ---
	%\bibliographystyle{apsrev4-2}
	%\bibliography{references2} 

	%\bibliography{references} 
	%\input{md1.bbl}
	
	\appendix*
	
	% Force section numbering back on for the appendix
	%\setcounter{section}{0}

	\section{Characteristic  function of p-RTP}
	\label{app:derivation}
	
	We first study the persistent Run-and-Tumble particle (p-RTP) subject to a constant background drift $v_d$.
	This system maps to the Hamiltonian with $\hat  p=i\partial_x$:
	\begin{equation}
		H_0 = i(v_d\sI + v\sz)\hat p + \lambda(\sI - \sx) - D \hat p^2 \sI \label{eq:1DCHam}
	\end{equation}
	Defining the state vector as $\ket{\mathbf{p}} = \begin{pmatrix}
		p_+\\p_-
	\end{pmatrix}$,
	the corresponding Fokker-Planck equation is diagonal in their advective fluxes:
	\begin{equation}
		\p_t \ket{\mathbf{p}}  = H_0\ket{\mathbf{p}}
	\end{equation}

	We begin with the two steady-state ODEs and the boundary conditions.
	First, we sum the two ODEs:
	With 
	$P(x) = P_+(x) + P_-(x)$ and $M(x) = P_+(x) - P_-(x)$, we have
	\begin{equation}
		D \frac{dP}{dx} - v_d P - v M = \text{const} \label{eq:app_j_const}
	\end{equation}
	The total current is $J(x) = J_+(x) + J_-(x)$. 
	Comparing this with Eq. \eqref{eq:app_j_const}, we see $\partial_x{J} = 0$. The boundary conditions statisfy
	\begin{equation}
		J(x) \equiv 0 \quad \forall x \in [-\ell, \ell]
	\end{equation}
	This gives the crucial algebraic relation:
	\begin{equation}
		v_d P(x) + v M(x) - D P'(x) = 0
	\end{equation}
	which we solve for $M(x)$:
	\begin{equation}
		M(x) = \frac{D}{v} P'(x) - \frac{v_d}{v} P(x) \label{eq:app_M}
	\end{equation}
	
	To get a single ODE for $P(x)$, we differentiate Eq. \eqref{eq:app_M}:
	\begin{equation}
		M'(x) = \frac{D}{v} P''(x) - \frac{v_d}{v} P'(x)
		M''(x) = \frac{D}{v} P'''(x) - \frac{v_d}{v} P''(x) \label{eq:app_Mpp}
	\end{equation}
	
	Aftersome straightforward manipulation we get the following ODE, which gives the characteristic in main text
	\begin{equation}
		D^2 P'''(x) - 2D\mu P''(x) 
		+ (v_d^2 - v^2 - 2\lambda D) P'(x) + 2\lambda v_d P(x) = 0
		\label{eq:app_ode_final}
	\end{equation}

	\subsection{Standard RTP:($D \to 0$)}
	
	For \(D = 0\), the system reduces and
	the stationary solution satisfies:
	\[
	\frac{d}{dx}[(v_0+v_d) p_+] = -\lambda p_+ + \lambda p_-, 
	\frac{d}{dx}[(v_0-v_d) p_-] = \lambda p_+ - \lambda p_-.
	\]
	Adding these equations gives zero total current and this integrates to:
	\[
	p_+(x) = A e^{r_1 x}, \quad \text{where} \quad r_1 = \frac{2\gamma v_0}{v_d^2 - v_0^2}.
	\]
	The total density is:
	\[
	P(x) = p_+ + p_- = A e^{r_1 x} \left(1 - \frac{v_d+v}{v_d-v}\right) = \frac{2v}{v - v_d} A e^{r_1 x}.
	\]
	Thus, the bulk solution is \(P(x) = C_1 e^{r_1 x}\). 
	
	At the boundaries, particles accumulate due to reflection, forming delta functions. The boundary masses are determined by balance between incoming flux and tumbling:
	\[
	(v_d+v) p_+(\ell) = \gamma m_+, \quad (v_d-v) p_-(-\ell) = \gamma m_-.
	\]
	Using \(p_+(\ell) = A e^{r_1 \ell}\) and \(p_-(-\ell) = -\frac{v_d+v}{v_d-v} A e^{-r_1 \ell}\), we find:
	\[
	m_+ = \frac{v_d+v}{\gamma} A e^{r_1 \ell}, \quad m_- = \frac{v_d+v}{\gamma} A e^{-r_1 \ell}.
	\]
	Normalization requires:
	\[
	\int_{-\ell}^{\ell} C_1 e^{r_1 x}  dx + m_+ + m_- = 1.
	\]
	Solving for \(C_1\) and \(A\), we obtain:
	\[
	C_1 = \frac{r_1}{2} \left[ \sinh(r_1 \ell) + \frac{v_d}{v} \cosh(r_1 \ell) \right]^{-1}, \quad
	A = \frac{v - v_d}{2v} C_1.
	\]
	The masses become:
	\[
	m_\pm = \frac{1}{2} \frac{v_d}{v} \frac{e^{\pm r_1 \ell}}{\sinh(r_1 \ell) + \frac{v_d}{v} \cosh(r_1 \ell)}
	\]
	Thus, the full solution is:
	\[
	{
		P_C(x) = \frac{r_1}{2} \frac{e^{r_1 x}}{\sinh(r_1 \ell) + \frac{v_d}{v} \cosh(r_1 \ell)} + m_\pm \delta(x \pm \ell) 
	}
	\]
	
	\subsection{$D \to \infty$ Limit}
	
	As \(D \to \infty\), the system reduces to drift-diffusion with drift \(v_d\):
	\[
	\frac{P'(x)}{P(x)} \approx \frac{v_d}{D}.
	\]
	The solution is:
	\[
	P_C(x) \approx \frac{v_d / (2D)}{\sinh(v_d \ell / D)} \exp\left( \frac{v_d}{D} x \right).
	\]
	Expanding for small \(v_d \ell / D\):
	\[
	{
		P_C(x) \approx \frac{1}{2\ell} \left( 1 + \frac{v_d}{D} x \right).
	}
	\]
	
	\subsection{Expressions for biased quantum trajectory}
	In small diffusion Limit ($D \to 0$), the distribution is defined on the domain $X \in [-\epsilon,\epsilon]$.
	$$
	P(X) = \mathcal{A} \frac{(1+X)^{\frac{r_1}{2} - 1}}{(1-X)^{\frac{r_1}{2} + 1}} + m_{\delta}^\pm \delta(X -\pm \epsilon) 
	$$
	where,
	$$
	r_1 = \frac{4\lambda J}{\gamma'^2 - 4J^2}
	$$
	$$
	\mathcal{A} = \frac{r_1/2}{\sinh(r_1 \ell) + \frac{v_0}{v}\cosh(r_1 \ell)}
	$$
	$$
	m_{\delta}^{\pm} = \frac{1}{2} \frac{v_0}{v} \frac{e^{\pm r_1 \ell}}{\sinh(r_1 \ell) + \frac{v_0}{v}\cosh(r_1 \ell)}
	$$
	In large diffusion Limit ($D \to \infty$), the asymptotic distribution for large diffusion (small Péclet number) is:
	$$
	P(X) \approx \frac{1}{2\ell} \left[ 1 - \frac{\gamma'}{\gamma_{eff}} \operatorname{arctanh}(X) \right] \frac{1}{1-X^2}
	$$
	\section{Tilted Belavkin--Barchielli Dynamics} \label{app:tBBZ}
	
	Under the reference measure $\Prob$, the unnormalized conditional state
	$\check{\sdm}_t$ obeys the linear Zakai equation
	\begin{equation}
		\dx \check{\sdm}_t
		= \Lop_{\mathrm{eff}}[\check{\sdm}_t]\,\dx t
		+ \sqrt{2\gamma}\,\Mop[\hat{L}]\,\check{\sdm}_t\,\dx Y_t ,
	\end{equation}
	where
	\begin{equation}
		\Lop_{\mathrm{eff}}[\sdm]
		= -i[\hat{H},\sdm]
		+ \gamma\bigl(2\hat{L}\sdm\hat{L}^\dagger
		- \{\hat{L}^\dagger\hat{L},\sdm\}\bigr),
		\qquad
		\Mop[\hat{L}]\sdm
		= \hat{L}\sdm + \sdm\hat{L}^\dagger ,
	\end{equation}
	and $Y_t$ is a Wiener process.
	
	To bias the measurement record, we introduce the exponential martingale
	$M_t^{(s)}=\exp(sY_t-\tfrac12 s^2 t)$ and define the tilted measure
	$\dx\Prob^{(s)}=M_t^{(s)}\dx\Prob$.
	The tilted unnormalized state using Ito calculus,
	$\widetilde{\sdm}_t=M_t^{(s)}\check{\sdm}_t$
	then satisfies
	\begin{equation}
		\dx \widetilde{\sdm}_t
		= \Bigl(\Lop_{\mathrm{eff}}
		+ s\sqrt{2\gamma}\,\Mop[\hat{L}]\Bigr)\widetilde{\sdm}_t\,\dx t
		+ \Bigl(\sqrt{2\gamma}\,\Mop[\hat{L}] + s\Bigr)\widetilde{\sdm}_t\,\dx Y_t .
	\end{equation}
	
	Writing $\hat{\sdm}_t=\widetilde{\sdm}_t/\Tr(\widetilde{\sdm}_t)$ and defining the quadrature expectation $\langle \hat{X}_L\rangle_t = \Tr(\Mop[\hat{L}]\hat{\sdm}_t)$ and the centered superoperator $\Hop[\hat{L}]\sdm = \Mop[\hat{L}]\sdm - \Tr(\Mop[\hat{L}]\sdm)\sdm$, one obtains, after normalization,
	\begin{equation}
		\dx \hat{\sdm}_t
		= \Lop_{\mathrm{eff}}[\hat{\sdm}_t]\,\dx t
		+ \sqrt{2\gamma}\,\Hop[\hat{L}]\,\hat{\sdm}_t\,\dx Y_t
		- 2\gamma\langle \hat{X}_L\rangle_t
		\,\Hop[\hat{L}]\,\hat{\sdm}_t\,\dx t .
	\end{equation}
	
	By Girsanov’s theorem, under $\Prob^{(s)}$ the innovation process
	\begin{equation}
		\dx \widetilde{W}_t
		= \dx Y_t
		- \bigl(\sqrt{2\gamma}\langle \hat{L}\rangle_t + s\bigr)\dx t
	\end{equation}
	is a standard Wiener process. The normalized dynamics therefore takes the
	compact form
	\begin{equation}
		\dx \hat{\sdm}_t
		= \Lop_{\mathrm{eff}}[\hat{\sdm}_t]\,\dx t
		+ \sqrt{2\gamma}\,\Hop[\hat{L}]\,\hat{\sdm}_t
		\bigl(\dx \widetilde{W}_t + s\,\dx t\bigr)
	\end{equation}
	with the associated signal equation
	\begin{equation}
		\dx Y_t
		= \bigl(\sqrt{2\gamma}\langle \hat{L}\rangle_t + s\bigr)\dx t
		+ \dx \widetilde{W}_t .
	\end{equation}

\end{document}